\documentclass[floats,floatfix,superscriptaddress,preprint]{revtex4-1}

\usepackage[pdftex]{graphicx}
\usepackage{amsmath}
\usepackage{amsfonts}
\usepackage{mathrsfs}
\usepackage[utf8]{inputenc}
\usepackage[T1]{fontenc}
\usepackage{color}
\usepackage{xcolor}
  
\renewcommand\vec{\boldsymbol}
\newcommand{\quotes}[1]{``#1''}

\begin{document}

\title{Topology of three-dimensional active nematic turbulence confined to droplets}


\author{Simon Čopar}
\thanks{These authors contributed equally to this work.}
\affiliation{Faculty of Mathematics and Physics, University of
  Ljubljana, Jadranska 19, 1000 Ljubljana, Slovenia}

\author{Jure Aplinc}
\thanks{These authors contributed equally to this work.}
\affiliation{Faculty of Mathematics and Physics, University of
  Ljubljana, Jadranska 19, 1000 Ljubljana, Slovenia}

\author{Žiga Kos}
\thanks{These authors contributed equally to this work.}
\affiliation{Faculty of Mathematics and Physics, University of
  Ljubljana, Jadranska 19, 1000 Ljubljana, Slovenia}
  
\author{Slobodan Žumer}
\affiliation{Faculty of Mathematics and Physics, University of
  Ljubljana, Jadranska 19, 1000 Ljubljana, Slovenia}

\affiliation{J. Stefan Institute, Jamova 39, 1000 Ljubljana,
  Slovenia}

\author{Miha Ravnik}

\affiliation{Faculty of Mathematics and Physics, University of
  Ljubljana, Jadranska 19, 1000 Ljubljana, Slovenia}

\affiliation{J. Stefan Institute, Jamova 39, 1000 Ljubljana,
  Slovenia}

\date{\today}

\begin{abstract}
Active nematics contain topological defects which under sufficient activity move, create and annihilate in a chaotic quasi-steady state, called active turbulence. However, understanding  active defects under confinement  is an open challenge, especially in three-dimensions. Here, we demonstrate the topology of three-dimensional active nematic turbulence under the spherical confinement, using numerical modelling. In such spherical droplets, we show the three-dimensional structure of the topological defects, which due to closed confinement emerge in the form of closed loops or surface-to-surface spanning line segments. In the turbulent regime, the defects are shown to be strongly spatially and time varying, with ongoing transformations between positive winding, negative winding and twisted profiles, and with defect loops of zero and non-zero topological charge.  The timeline of the active turbulence is characterised by four types of bulk topology-linked events --- breakup, annihilation, coalescence and cross-over of the defects --- which we discuss could be used for the analysis of the active turbulence in different three-dimensional geometries. The turbulent regime is separated by a first order structural transition from a low activity regime of a steady-state vortex structure and an offset single point defect. We also demonstrate coupling of surface and bulk topological defect dynamics by changing from strong perpendicular to inplane surface alignment. More generally, this work is aimed to provide insight into three-dimensional active turbulence, distinctly from the perspective of the topology of the emergent three-dimensional topological defects. 
\end{abstract}

\maketitle
\section{Introduction}
The ability of the material to employ externally or internally stored energy to spontaneously organize, flow, move, or change shape --- i.e. be active --- is found in a variety of materials~\cite{MarchettiMC_RevModPhys85_2013}. A major class of active materials are active nematics~\cite{DoostmohammadiA_NatCommun9_2018}, which exhibit strong collective behaviour and self organisation, emergent as local orientational order of material building blocks in  material systems such as bacteria~\cite{BlanchMercaderC_PhysRevLett120_2018} and microtubule-kinesin mixtures~\cite{SanchezT_Nature491_2012,NeedlemanD_NatRevMater2_2017}. Active nematic formalism could be also applied to describe other biological systems, such as cytoskeleton~\cite{BruguesJ_ProcNatlAcadSci111_2014,ProstJ_NaturePhys11_2015} and biofilm dynamics~\cite{SawTB_Nature544_2017,KawaguchiK_Nature545_2017}. 
A recurring behaviour in active nematics is the formation of topological defects, i.e. regions of broken orientational order, which are subject to topological invariant conservation~\cite{AlexanderGP_RevModPhys84_2012}. Due to strong localized deformation of the orientational order, some active nematic defects can perform as strong effective sources of material flow, as  dependent on the symmetry of the defects  \cite{GiomiL_PhysRevLett110_2013,DeCampSJ_NatMater14_2015}. A recurring property of topological defects in different active nematic systems is that at larger activities, they undergo into a regime of irregular and chaotic motion at low Reynolds numbers --- called active turbulence. In the turbulent regime, generally, the topological defects are constantly created and annihilated~\cite{SanchezT_Nature491_2012,BlanchMercaderC_PhysRevLett120_2018} which can be characterised  by using statistical tools from classical turbulence~\cite{GiomiL_PhysRevX5_2015,UrzayJ_JFluidMech822_2017}. The threshold between regular and irregular dynamics is strongly dependent on the confinement~\cite{HenkesS_PhysRevE97_2018,DoostmohammadiA_NatCommun8_2017}, higher order force multipoles~\cite{MaitraA_ProcNatlAcadSci115_2018} and friction, which can even lead to stabilization of ordered defect phases in active nematics ~\cite{DeCampSJ_NatMater14_2015,PutzigE_SoftMatter12_2016}. 


Geometrical confinements is today seen as one of the prime mechanisms for controlling the dynamics of active nematics~\cite{WuK_Science355_2017,WiolandH_NaturePhys12_2016,LushiE_ProcNatlAcadSci111_2014,TheillardM_SoftMatter13_2017}, notably also in the context of possible energy harvesting~\cite{ThampiSP_ScienceAdvances2_2016}. Active nematics interact with their confinement  through hydrodynamic boundary conditions, such as slip, no-slip and friction, as well as  through boundary conditions on the surface imposed orientational ordering, also called surface anchoring. Active nematic droplets based on the mixture of microtubules and molecular motors were shown to periodically morph their shape~\cite{KeberFC_Science345_2014}. Active nematic layers organised at the surface of spherical droplets are also explored from the perspective of  topological defect trajectories and their mutual interactions~\cite{KeberFC_Science345_2014,KhoromskaiaD_NewJPhys19_2017,ZhangR_NatCommun7_2016,
HenkesS_PhysRevE97_2018}. When interfacing an ordered passive nematic fluid, active nematics can also exhibit directional streaming along the passive nematic director \cite{GuillamatP_ProcNatlAcadSci113_2016}, and actuate the defects in the passive nematic host  \cite{GuillamatP_ScienceAdvances4_2018}. Fluid flow inside a droplet can produce net propulsion of the droplet, which itself acts as an individual active swimmer particle~\cite{MaassCC_AnnuRevCondensMatterPhys7_2016,GaoT_PhysRevLett119_2017,TjhungE_ProcNatlAcadSci109_2012}.

Realising and controlling three-dimensional active nematic materials is an open emergent challenge in experiments~\cite{OpathalageA_ProcNatlAcadSci116_2019,ZhouS_NewJPhys19_2017,NishiguchiD_NatCommun9_2018,beller_talk,ScholichA_arXiv190408886,dogic_talk} and numerical simulations~\cite{UrzayJ_JFluidMech822_2017,WhitfieldCA_EurPhysJE40_2017,ShendrukTN_PhysRevE98_2018,
HartmannR_NaturePhys15_2019}, where one of the standing major complexities is how to characterise the complex 3D spatially and time-varying flow and orientational fields, that moreover have embedded topological defect regions~\cite{UrzayJ_JFluidMech822_2017}.
Analogous systems exist that are dominated by topological defects in three-dimensions, distinctly  passive (non-active) liquid crystal colloids~\cite{TkalecU_Science333_2011,MartinezA_NatMater13_2014} and chiral ferromagnets~\cite{PfleidererC_Nature465_2010,FosterD_NaturePhys}, where structures are controlled and characterised by using topological approaches. Indeed, due to the apolar nature of nematic director, the nematic (passive or active) in three dimensions can form not only point defects (as in 2D), but also defect lines and loops, which can be characterised with different topological invariants, including winding number, topological charge and self linking number~\cite{AlexanderGP_RevModPhys84_2012,CoparS_PhysRep538_2014}. In passive nematics, coarsening structures during relaxation of a quenched state from isotropic to nematic also show notable analogy in view of defect loop topology to 3D active turbulent defect profiles~\cite{NikkhouM_NaturePhys11_2015}. To generalize, the idea of this paper is to apply concepts of topology to characterize structural events in 3D active nematic turbulence.

In this paper, we explore the topological defects regimes of three-dimensional active nematic under the  spherical confinement and no-slip surface, specifically focusing on the topology-affecting events in the active turbulence. We show that for moderate activities, the active nematic assumes an oriented state with an off-center shifted single point defect in the form of a small loop, and vortex flow with direction-set angular momentum. Upon increasing activity, we observe the onset of topological turbulence, characterised by a first-order structural transition and hysteresis between the offset point defect and the turbulent regime, also showing the corresponding phase diagram as dependent on activity and droplet size. In the turbulent regime, we show that the active turbulent dynamics can be interpreted as a time-series of topological defect affecting events: defect splitting, annihilation, merging and crossovers. Depending on surface anchoring conditions, the defects are in the form of closed loops (perpendicular anchoring) or surface-to-surface lines that effectively wet the  confining surface (inplane anchoring), resulting in interesting surface-to-bulk conditioned turbulent dynamics. More generally, the results demonstrate dynamics of strongly confined active nematic and point towards using concepts of topology  for controlling the three-dimensional active nematic turbulence.

\section{Results}
The defect phenomena of the active nematic is explored in an elementary confinement of a droplet with fixed spherical shape by using mescoscopic numerical modelling of active nematodynamics~\cite{DoostmohammadiA_NatCommun9_2018, HatwalneY_PhysRevLett92_2004, GuillamatP_ScienceAdvances4_2018,GiomiL_PhysRevLett110_2013}, which was shown to give good agreement with experiments on dense active nematic systems~\cite{Wensink_Yeomans_PNAS_2012,DoostmohammadiA_NatCommun9_2018,GiomiL_PhysRevX5_2015}. The approach relies on the dynamic coupling between the material flow and the mesocospic order parameter tensor $Q_{ij}$ that covers the orientational ordering of active nematic. The activity is described by dipolar-like forcing via the active stress contribution. The confining spherical surface is set to impose strong perpendicular (homeotropic; in Figs. 1--3) or inplane (degenerate planar; in Fig. 4) alignment of the active nematic at the surface, which experimentally, would correspond to different surface functionalities~\cite{MartinezPratB_NaturePhys15_2019}. Using strong anchoring and fixed spherical shape  allows us to discern the effects of shape from the effects of topology that are otherwise inherently intertwined. Generally, such fixed shape regime corresponds to having materials with large surface tension or a background medium that is rigid enough to support the shape of the active nematic droplet (e.g. gel-like \cite{EllisPW_NaturePhys14_2018}). We use no-slip boundary condition at the surfaces to simulate the host medium that resists flow. All distances are measured in units of the nematic correlation length $\xi_N$, time is measured in units of the intrinsic nematic time scale $\tau_\text{N}$, and activity $\zeta$ in the units of $L/\xi_N^2$ where $L$ is the single elastic constant of the material. Equally, activity can be also described with the active length $\xi_\zeta=\sqrt{L/\zeta}$. Note that the nematic correlation length measures the effective size (thickness) of the defects  and is given as a relative strength of the nematic elasticity vs. variations in nematic order, which are --- beside the activity and confinement --- the key energetic mechanisms that affect the formation, structure and dynamics of topological defects. For more on the approach, please see Methods. Finally, note that this work focuses on the defect phenomena in three-dimensional active nematic, i.e. active nematic in the whole bulk of the spherical droplet, which is different to many current works which consider a thin layer of active nematic material, for example at the surface of a droplet~\cite{ZhangR_NatCommun7_2016, GuillamatP_ScienceAdvances4_2018}

\subsection{Active regimes in spherical droplet}

\begin{figure*}
  \includegraphics[width=0.8\textwidth]{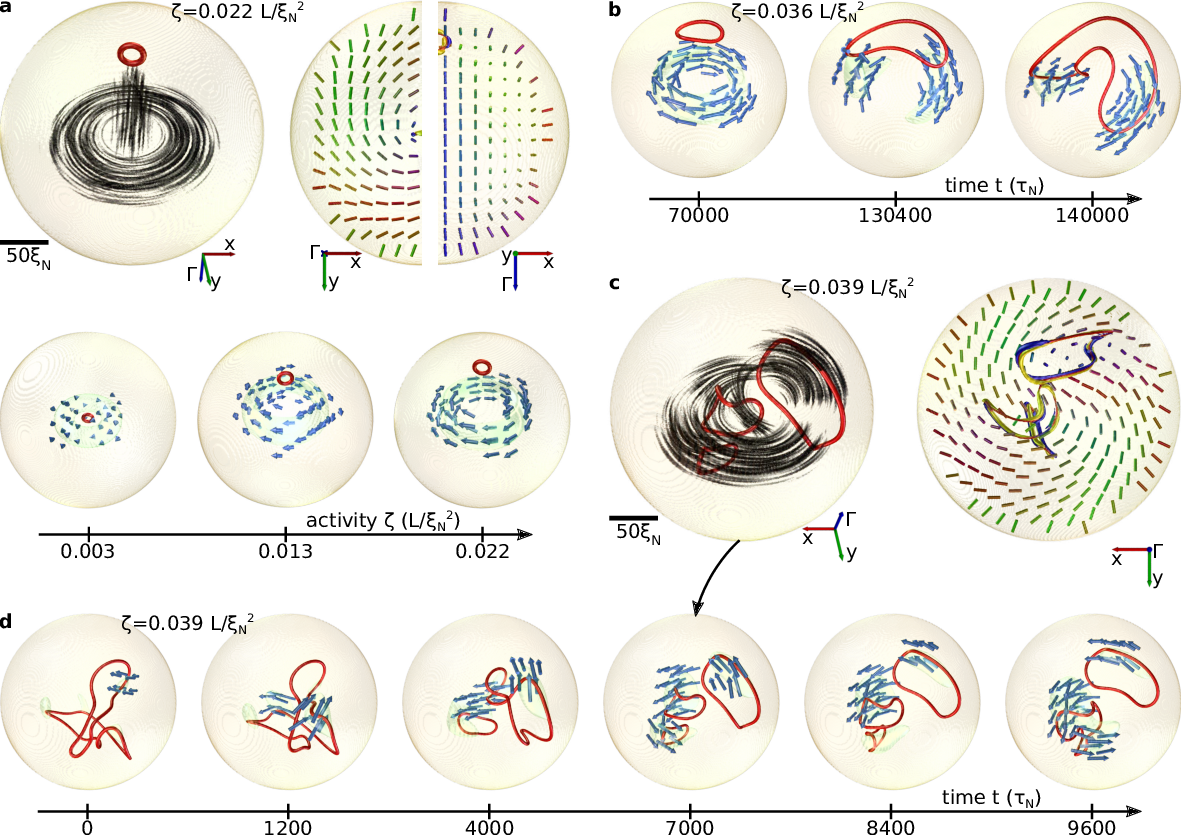}
  \caption{
    \textbf{Active nematic regimes in spherical droplet confinement with perpendicular surface alignment.}
    Snapshots of  dynamics for different activities, showing time evolution, velocity streamlines and director field. (a) Active nematic regime at low activities with single point defect (in red, in the form of a small ring) displaced from the droplet center and flow vortex (gray streamlines) with the angular momentum vector $\boldsymbol{\Gamma}$. Note that the flow velocity component in the direction of $-\vec{\Gamma}$ is $\sim 2.5$-times smaller than the flow inplane component in the $xy$ plane. Right panel shows director profiles in the plane of the vortex and perpendicular to the vortex. Red color corresponds to the director along $x$ axis, green color along $y$ axis and blue color along $\vec\Gamma$. Bottom panel shows how with increasing activity the point defect (in red) moves further away from the droplet center and flow magnitude (blue arrows) increases. (b) Transition of point defect regime into the turbulent regime upon an increase of activity from $0.035\,L/\xi_\text{N}^2$ to $0.036\,L/\xi_\text{N}^2$; note stretching and  deformation of the defect loop, and the disintegration of the regular velocity profile. (c) Turbulent active nematic regime. Defect loops are shown as isosurfaces of reduced degree of order (in red) and flow field with gray streamlines. Right panel shows the corresponding director in the given plane; the local variability of the defect cross section is illustrated  with splay-bend parameter (yellow and blue isosurfaces; see Methods). (d) Selected time-line of active turbulence in the active nematic droplet. Red isosurface of the scalar order parameter is drawn at $S=0.45$; blue arrows and bright green isosurfaces represent velocity field above selected cut-off magnitudes (see Methods).
  }
  \label{fig:fig1}
\end{figure*}

\begin{figure*}
  \includegraphics[width=\textwidth]{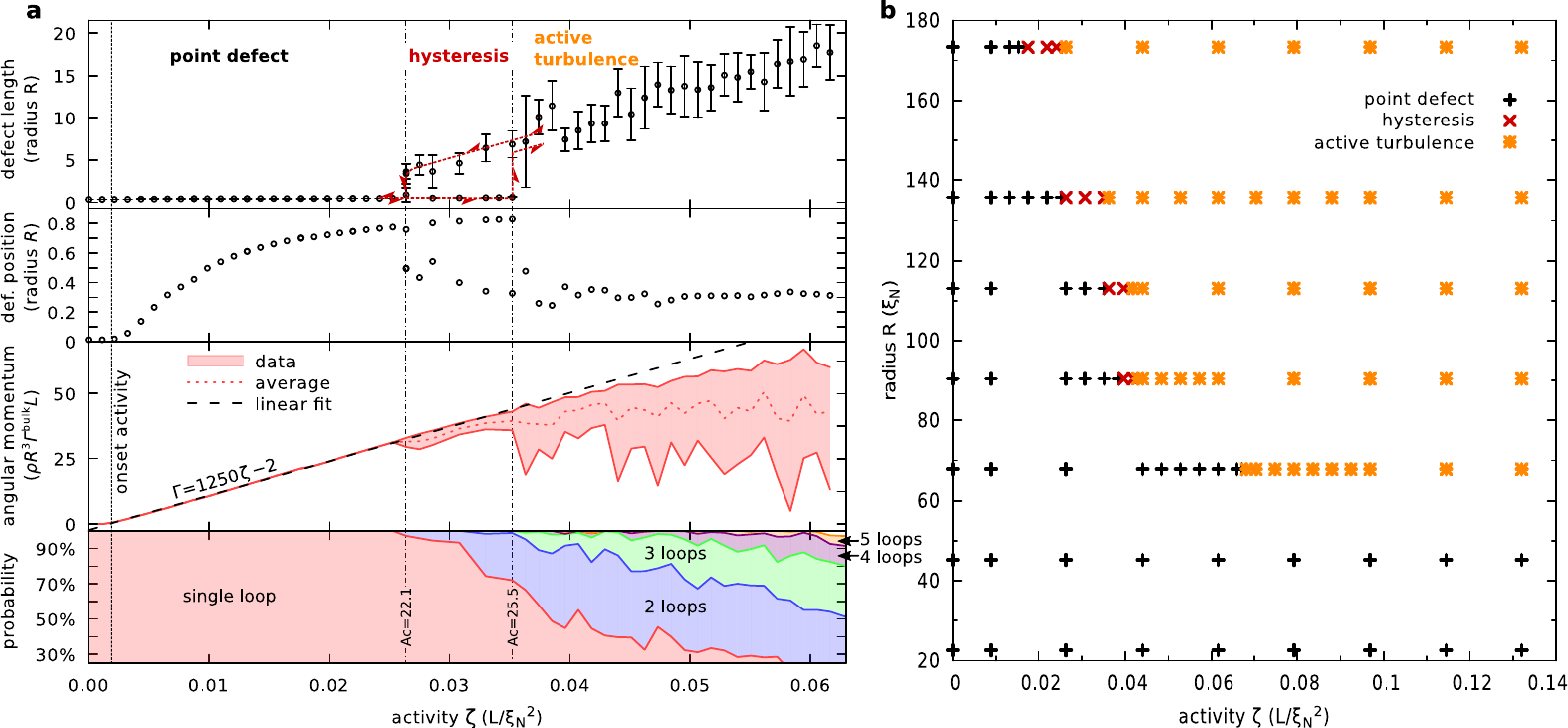}
  \caption{
    \textbf{Analysis of active nematic regimes in  confinement of spherical droplet.}
    (a) Multiple material characteristics of the three-dimensional active nematic in spherical droplet confinement of radius $R=136\xi_\text{N}$ with perpendicular surface alignment as dependent on the activity. The calculations are performed for increasing or decreasing activity, clearly showing a pronounced hysteresis. At each activity change, previous structure is taken as the initial condition and simulation is run for $23000\,\tau_\text{N}$ for each data point. Top panel shows the total length of defect loop cores at a given activity, with the bars indicating the standard deviation of the natural oscillation of the defect length over time. Second panel in (a) shows average radial distance of the defect core regions from the droplet center. Third panel gives the total angular momentum set by the flow and its standard deviation in time. The bottom panel gives the probability for finding given number of defect loops within the droplet.  Note three active regimes  --- point defect, hysteretic regime and developed active turbulence --- each with different response of the active material characteristics. Linear fit of the angular moment dependence on the activity, in the point defect regime, is used to set the onset activity below which the vortex flow profile has negligible magnitude and the point defect remains at the droplet center. 
    Graphs of angular momentum and probability are shown only for hysteresis behaviour at decreasing activity.  Hysteretic region is delimited by activity numbers $Ac\approx 22.1$ and $Ac\approx 25.5$, for the considered droplet size. (b) Phase diagram of active regimes in spherical droplets with perpendicular surface alignment, as given by droplet radius $R$ and activity $\zeta$. }
  \label{fig:fig2}
\end{figure*}

Depending on the activity of the material, two active regimes with distinct behaviour of topological defects are found for the three-dimensional bulk active nematic in the confinement of a spherical droplet with perpendicular alignment: for low activity, 
structure with an offset stationary point defect is observed (Fig.~\ref{fig:fig1}a) whereas for higher activities, a regime of three-dimensional active turbulence with spatially- and time-varying defect loops is observed (Fig.~\ref{fig:fig1}c). These two regimes are separated by a (hysteretic) structural transition (Fig.~\ref{fig:fig1}b). In the limit of no activity (i.e. also no material flow), the considered spherical confinement would
exhibit a radial nematic director structure with a single $+1$ radial hedgehog point defect at the center, as imposed by the confining surface conditions; indeed, such profile is observed for activities below the activity $\zeta_\text{onset}=1.6\times 10^{-3}\,L/\xi_\text{n}^2$ (see also Fig. 2).
Note that the hedgehog point defect appears in the form of a small ring of local winding number $+1/2$, with diameter of the order of nematic correlation length, and is topologically equivalent to a point~\cite{WangX_PhysRevLett116_2016}; the exact small-scale structure will in practice depend on the microscopic building blocks of the material. Therefore, in view of our work, loops with radius of the order of the nematic correlation length will be called point defects. Above the activity $\zeta_\text{onset}$ a self-sustained structure of a twisted director profile and a vortex flow appears, characterised by the orientation and magnitude  of the angular momentum vector (Fig.~\ref{fig:fig1}a).  Similar as in (passive) cholesteric nematic droplets~\cite{PosnjakG_NatCommun8_2017}, twisted director field pushes the point defect away from the droplet center. Since our system has no preferred chirality, the shift of the point defect can be both along or against the angular momentum vector. Additionally, there is a velocity field component of roughly $\sim 2.5$-times smaller magnitude then the circular vortex flow which goes along the center of the droplet towards  the point defect and then near the surface back to the opposite end of the spherical droplet confinement.
For increasing activity, the magnitude of the vortex flow increases and the defect shifts more and more toward the surface of the droplet, and as a critical activity is reached, the small defect ring opens into a larger loop (see Fig.~\ref{fig:fig1}b). At this value of the activity, the activity-induced flow can overcome the elastic barrier of the nematic field  to transform the point defect into a topologically equivalent large defect loop.  This process involves contributions of defect core line tension, nematic elasticity, advection, backflow, shear active flow as well as coupling to confining geometry and actual local and global defect topology. Above the critical activity, the active nematic transitions into a chaotic irregular behaviour --- i.e. the active turbulence regime. There can be strong material flow due to large elastic deformation around the core of the loop, and overall, strong dynamic deformations of the loop are observed, including topology-modifying events, such as breakup of a loop into multiple loops, as shown in Fig.~\ref{fig:fig1}c,d.

Performing a more quantitative study of the active nematic regimes in the spherical droplet confinement, several distinct material characteristics are analysed (as shown in Fig.~\ref{fig:fig2}), that  can assess the  three-dimensional nematic orientational ordering and material flow, as well as the topology of the defects. For each simulation, we changed the activity between two  neighbouring data points in Fig.~\ref{fig:fig2}a, let the system evolve for $4500\,\tau_\text{N}$ to reach a dynamic steady state, and then over a time interval of $18500\,\tau_\text{N}$ collected data on total defect length, (average) defect position, total angular momentum, and probability for observing N-number of defect loops (Fig.~\ref{fig:fig2}a). Note that, in  the analysis we employ direct 3D tracking of defect line cores based on variations in the nematic degree of order, as further explained in Methods. The defect length inside the droplets quantifies that at low activities there is a regime of a single point defect, but upon increasing activity, a structural transition occurs to the active turbulent regime, in which the single defect loop or multiple loops dynamically transform and vary in-time, with total defect line length changing in time at the order of multiple $10\%$ from the average value. The position of the centre of the point defect changes with increasing activity and shifts from the droplet centre towards the surface of the droplet. 
As the active nematic transitions  to the turbulent regime, the active defect loops explore and transverse the available volume  irrespective of actual activity. 
The average radial position of defect segments from the droplet center settles in the turbulent regime to $\approx 0.3\,R$, as at high activities the defect segments become more uniformly distributed. The average defect position also shows hysteretic behaviour.  The average defect core position from the droplet center well measures the point defect position in the low activity state, but in the turbulent regime, it is more meaningful to observe the probability density of defect line segments per unit volume, with respect to relative distance from the center of the droplet, as shown in Supplementary Fig. 1.
The angular momentum grows within the point defect regime continuously with activity, according to a linear dependence $(\zeta-\zeta_\text{onset})$. $\zeta_\text{onset}$ is the activity for the onset of vortex flow, which we determine numerically by a linear fit of the angular momentum dependence on activity. Note that such emergence of active flows in confined and defect systems is known to depend on the symmetry and profile of the equilibirum (no-activity) structure~\cite{VoituriezR_EurophysLett70_2005,GreenR_PhysRevFluids2_2017,Kruse_Sekimoto_PRL_2004}. When the active nematic passes to the active turbulence regime, the average magnitude of the angular momentum gradually saturates with activity, which again signifies  the break up of the regular defect dynamics.
At this transition between point defect and active turbulence regimes, also the process of defect loop breakup (i.e. creation of new loops) and annihilation of emerging loops begins, resulting in a growing fraction of time when there is more than one loop present in the spherical droplet.

Activity is the main control parameter for the transition between the point defect regime and the active turbulence; however, this transition could be also driven for example by changing  the active nematic elastic constant or the size of the confining cavity. Figure~\ref{fig:fig2}b presents a phase diagram of active regimes with respect to activity and droplet size, showing that larger droplets transition into active turbulent regime at lower activities, whereas smaller droplets require larger activites to undergo structural transition from regular to irregular dynamics. Indeed, such behaviour can be well understood by considering the dimensionless activity number $Ac=\sqrt{\zeta R^2/L}$ which gives the ratio between the confinement size $R$ and  active length $\xi_\zeta$. In Fig.~\ref{fig:fig2}b, the transition from point defect to active turbulence is observed roughly at $Ac\sim 20$.

The structural transition between the point defect structure and the active turbulence  exhibits a clear hysteresis, which is a signature of discontinuous (first-order) transition between the two structures (Fig.~\ref{fig:fig2}a). The ordered phase with the offset point defect can be ``over-activated'' (i.e. with activity above the transition value), which we simulate by gradually increasing activity in small steps from the low activity state until the turbulence appeared, assuring for each step in the activity change that the system reaches the dynamic steady state. Alternatively, the turbulent state can be ``under-activated'' (i.e. with activity below the transition value), before the turbulent defect loop collapses into a point defect, which we simulate by gradually decreasing activity in small steps. The turbulent state can only appear when the flow is strong enough to overcome the elastic tension force of the defect line. Once the defect line is extended, it can remain extended even at lower activities.
 In the turbulent regime the system behaves chaotically, where changes  in the shape of defects can lead to a very different future evolution of the system. The system has no  static equilibrium, but reaches a dynamical steady state in which macroscopic variables, such as the defect length, angular momentum and average number of loops settle to fluctuating around an average value.


\subsection{Topology of defect loops in active turbulence}

\begin{figure*}
  \includegraphics[width=\textwidth]{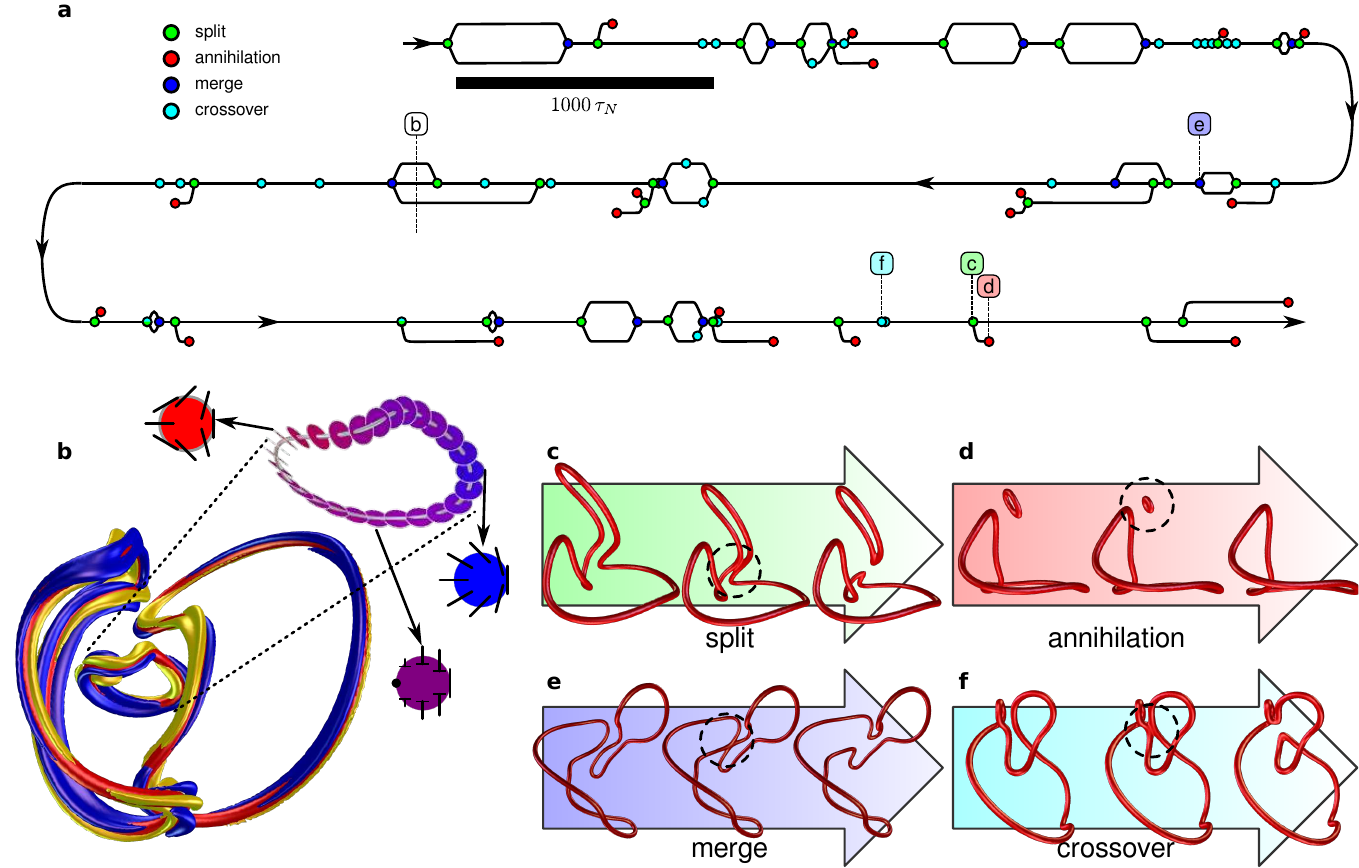}
  \caption{
    \textbf{Topological events in turbulent regime.}
    (a) Graphically depicted topological flow diagram for the active nematic defects loops in a selected time interval. Note that the diagram time axis is in scale (black line) and is given in units of the intrinsic nematic time scale $\tau_\text{N}$.  At all times, one of the loops carries topological charge $+1$ and other have zero charge. (b) A snapshot of active turbulence with three loops. The color-coding of the disks for the selected defect loop shows local winding number of the defect line, where blue refers to $+1/2$, red to $-1/2$ winding number and purple to local twist defect line profile. The variation of the defect line cross section is also illustrated more qualitatively with the splay-bend parameter (yellow and blue isosurfaces).
    (c-f) Time-sequences of selected loop topology-changing events: splitting, annihilation, merging and crossover. Panel (c) shows splitting of a neutral loop (upper) from a loop with a +1 topological charge (lower). The neutral loop soon annihilates in panel (d). Droplet radius is $R=136\,\xi_N$ and activity $\zeta=0.04\,L/\xi_\text{N}^2$.
  }
  \label{fig:fig3}
\end{figure*}

Topology is a natural tool for studying topological defects that goes beyond exact detailed material characteristics, but rather focuses on the overall structure, and we use topological analysis for the characterisation of active turbulent regime under the confinement of active nematic into spherical droplets.
We focus on topological properties of the actual defect loops --- their number, how they are connected to each other, and their topological charge --- as well as on the local variation of the nematic director surrounding the defect loops, i.e. considering defect loops not only as lines but rather as ribbon-like objects. Figure~\ref{fig:fig3} shows the dynamics of active nematic turbulence in the spherical droplet, analysed from the perspective of the topology of three-dimensional active defect loops. In Fig.~\ref{fig:fig3}a, a time series of defect topology-affecting events is presented --- importantly, with time-axis in scale --- for a selected time interval of the active tubulence in an active nematic droplet ($R=136\xi_N$ and $\zeta=0.04\,L/\xi_\text{N}^2$). Four different topological events are observed to occur stochastically in time, with the average event rate set by (and increasing with) the activity. New loops of zero topological charge are observed to detach (pinch-off) from a preexisting loop (the existence of at least one loop is topologically enforced by the droplet perpendicular surface conditions), extra loops annihilate, or merge back into the loop with nonzero charge, which by itself cannot disappear. Crossovers of the defect loop with itself can rewire the loop, reversing the path orientation of one part relative to the other. More generally, the visualised map reflects the conservation laws that would formally be described with a Morse theory, where topological events are critical points (saddles, endpoints) of the manifold comprised of the defect loci over time (the world sheet)~\cite{book-Milnor}. Examples of loop splitting, annihilation of a single loop, merger of two loops, and a crossover (i.e. self-intersection) are shown in snapshots in Fig.~\ref{fig:fig3}c-f  and in more detail in Supplementary Fig. 2.

The director cross section of defect loops in the regime of active turbulence varies along the defect loop in a complex way, and also changes in time, as shown in Fig.~\ref{fig:fig3}b and Supplementary Fig. 3. The defect loops in the three-dimensional active turbulent regime exhibit sections of local director symmetries of $+1/2$ and $-1/2$ winding number and twisted profile, which has profound effect also on the local dynamics and local active flow. It is well known from 2D active nematic systems that $+1/2$ defect profiles are generally strong generators of flow and strongly propel this type of (in 2D) point defects, whereas $-1/2$ defect profiles do not generate flow and are less motile~\cite{GiomiL_PhysRevLett110_2013}. Indeed, it is this variability of the director field that we now observe in the local cross-sections of the three-dimensional defect loops (see Supplementary Fig. 3), which then affects the dynamics of not only itself but all defect loops, through both nematic elasticity and hydrodynamics. The role of local cross-sections is known to affect the dynamics of reconnection events between defect line segments~\cite{HindmarshMB_RepProgPhys58_1995,IshikawaT_EurophysLett41_1998}. In our system, immediately before and after topological reconnection events the director orientations of both defect line segments generally lie in the same plane, and rewire in a way that makes the director between the defect lines uniform (without introducing topological solitons). For example, two defect line segments with locally twisted profile may reconnect into another pair of twisted profiles, a pair of $+1/2$ and $-1/2$ local profiles, or other intermediate profiles with a net neutral winding number (see Supplementary Fig. 2). Finally, the fact that the defect lines in spherical droplet confinement with perpendicular conditions are closed into loops imposes that this local director profile ($+1/2$, $-1/2$ or twisted) must come back to the same orientation when encircling the loop, which encodes the total topological charge carried by the loop~\cite{CoparS_PhysRep538_2014}.

At the level of the topology of the loop as a whole with surrounding director, we observe that one of the loops always carries a $+1$ topological charge --- set by the $+1$ topological charge imposed by the perpendicular (homeotropic) surfaces --- whereas the rest are topologically neutral (i.e. charge zero) and can annihilate without contacting another loop (see Fig.~\ref{fig:fig3}d), or alternatively, also merge back with another loop. From purely topological perspective, also pairs of oppositely charged loops could be formed (not observed in our simulations), especially in larger systems and with larger activities, while still preserving the net charge conservation set by confinement. However, the formation of oppositely charged pairs is --- at least in the considered material regime --- energetically even more unfavourable as the formation of zero-charged defect loops. Finally, a possible further approach to classify the active defect loops topologically could involve quaternionic approach \cite{CoparS_PhysRep538_2014} or construction of Pontryagin-Thom surfaces \cite{ChenBG_PhysRevLett110_2013}.

\subsection{Surface-contacting topological defects in active turbulence of spherical droplets with inplane alignment}

\begin{figure*}
  \centering
  \includegraphics[width=0.75\textwidth]{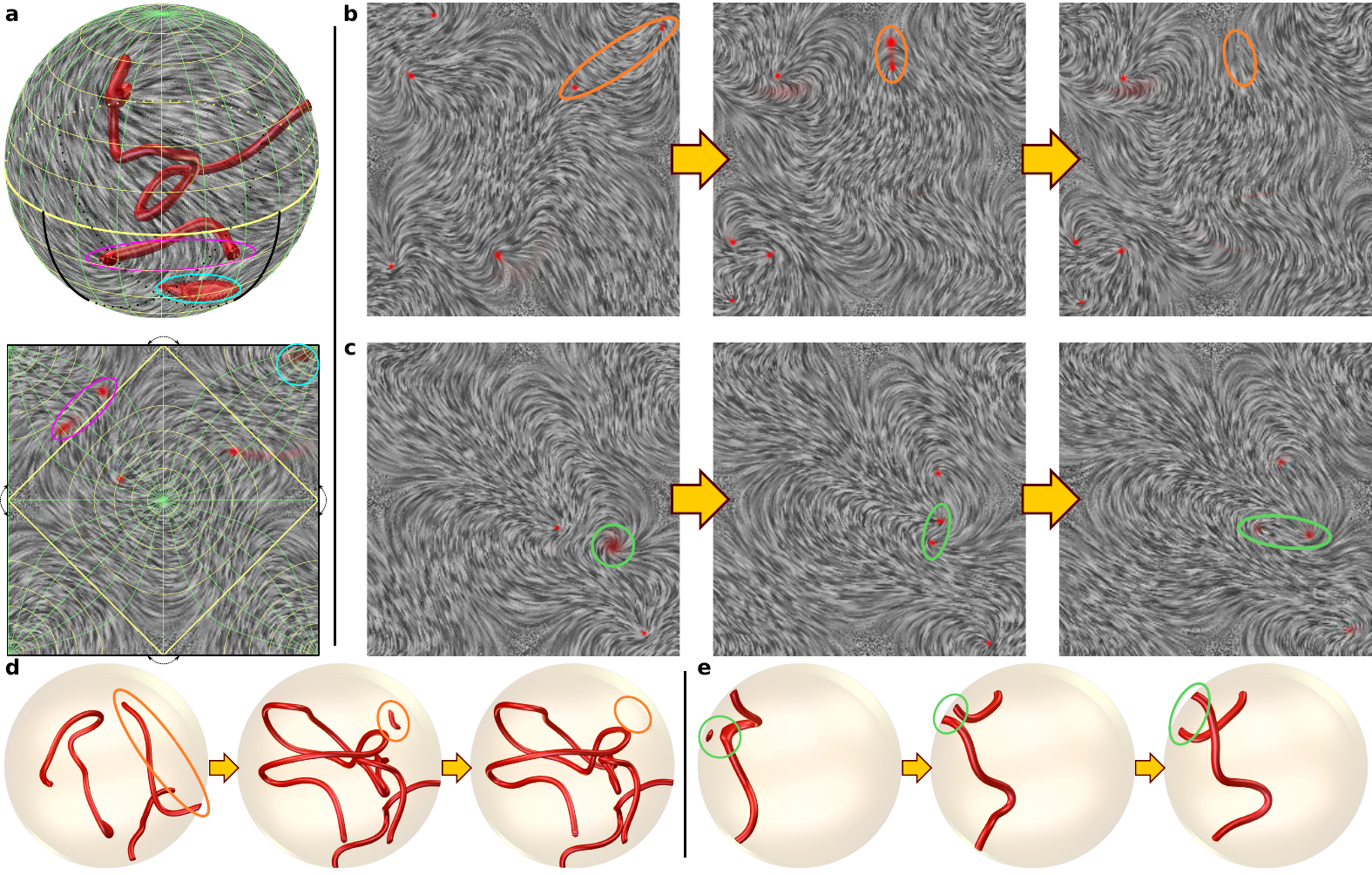}
  \caption{
    \textbf{Surface and bulk active nematic topological defect transitions in spherical  droplets with planar degenerate surface alignment.}
    (a) Snapshot of active nematic droplet showing two bulk defect lines, which terminate at the surface (ends of one are encircled with magenta), and a boojum defect (encircled in cyan). The inplane director field at the spherical droplet surface is represented by grayscale streaks.
   Bottom panel shows Peirce quincuncial projection map of the surface director and defects from the entire sphere onto a square; the sphere can be reproduced by folding four corners of the square back together into one point where the yellow square represents the equator of the sphere. Note the pointed out defects encircled in magenta and cyan, as shown in the sphere and the projection square.       
    (b) Annihilation sequence of a selected $\pm 1/2$ defect pair at the surface of the active nematic droplet shown using Peirce projection map. During the sequence a surface profile with five $+1/2$ defects and one $-1/2$ defect is transformed into four $+1/2$ defect profile, conserving the $+2$ surface topological charge set by the Euler characteristic of the sphere. In the bulk of the droplet, the sequence corresponds to a line segment beeing annihilated on the surface, as shown in (d). The annihilating surface defect pair is encircled in orange.
    (c) At the droplet surface, $+1$ surface boojum breaks into two $+1/2$ defect profiles (encircled in green). In the bulk, this surface process corresponds to a defect line segment interacting with the surface boojum, resulting in two bulk defect lines, as shown in (e). 
    (d) Full view of the changes in the bulk defect structure (in red, drawn as isosurfaces of $S=0.51$) that corresponds to the defect annihilation shown in (b). 
    (e) Full view of the changes in the bulk defect structure (in red, drawn as isosurfaces of $S=0.523$) that corresponds to the breaking of a boojum  shown in (c)
  }
  \label{fig:planar}
\end{figure*}

Active turbulence in active droplets of fixed spherical geometry with strong surface degenerate inplane alignment (i.e. planar degenerate anchoring) is explored, showing a system of interesting surface-to-bulk coupled defect dynamics. Under degenerate planar anchoring, the defect lines within the bulk are not any more necessarily closed into loops, but can also terminate at the surface, as shown in Fig.~\ref{fig:planar}. Indeed, in addition to closed defect loops observed for droplets with perpendicular surface alignment, defect lines with surface-to-surface spanning ends and surface boojum defects are observed, resulting in the turbulent defect dynamics which in the bulk is generally similar to the dynamics in homeotropic droplets but it is additionally coupled to the topological events at the droplet surface.  Figure~\ref{fig:planar} shows two such surface events using the Peirce quincuncial projection of the surface director and defects from the sphere onto a square: in Figs.~\ref{fig:planar}b,d, a pair of $\pm 1/2$ defects within the surface director field is annihilated, and in Figs.~\ref{fig:planar}c,e,  a surface boojum is split into a pair of $+1/2$ defects.
During the pair surface annihilation event, a structure with four $+1/2$ defects and a $\pm 1/2$ defect pair is transformed into a structure with only four $+1/2$ defects. In the bulk, such event: (i) either corresponds to the annihilation of a defect surface-to-surface line segment on the surface of the droplet (which was the actual case in the presented simulation, as shown in Fig.~\ref{fig:planar}d), or (ii) a merger of two ends of defect line segments into one segment and the following detachment of the defect  segment from the surface. 
In Fig.~\ref{fig:planar}c, a $+1$ surface boojum interacts (and merges) with a section of a bulk defect, which results in splitting of the boojum into two $+1/2$ defects, transforming the surface director profile with two $+1/2$ defects and the $+1$ boojum into a surface field with four $+1/2$ defects. In the bulk of the droplet, the defect line that merged with the boojum splits into two parts, each of them terminating at one of the newly created $+1/2$  defects at the surface, as shown in Fig.~\ref{fig:planar}e. Additionally, not shown here, we observe creation of surface defect pairs when an existing bulk defect loop segment is pushed from the bulk to the surface, where it is split into two ends (seen at the surface as a pair of $\pm 1/2$ defects), or also alternatively, a reconfiguration of the surface charges by merging of combinations of $-1/2$ defects and $+1$ boojums.

Topologically, the observed active turbulent defect dynamics in droplets with inplane alignment is determined by the director on the surface where the total winding number (2D topological charge) must be equal to the Euler characteristic of the confining surface $\chi$.
Specifically, in our droplets, the net 2D surface topological surface charge must equal $\chi=2$, which in the limit of zero activity, is realized by a pair of $+1$ boojums at the opposite poles of the droplet. However, for non-zero activity, as shown if Fig.~\ref{fig:planar}, the surface boojums are only rarely observed and the surface charge is rather distributed between four $+1/2$ defects and additional pairs of $\pm 1/2$ defects, which are actually endings of the bulk $+1/2$ or $-1/2$ defect lines touching the droplet surface. More generally, the topological dynamics of the defect lines (and loops) in the 3D bulk of active nematic induces a lower dimensional active defect dynamics at the confining surfaces. 

The turbulent defect dynamics at the surface of the droplet shares similarities with the 2D active nematic turbulence (e.g. in films or shells); however, it is different from the point that it is actually coupled to the active flow and defect reconfiguration of the entire bulk. The no-slip boundary condition for the material flow, used in this work, effectively  prevents surface defects to be driven by the activity directly on the droplet surface, but it is the active dynamics of bulk defects that cause apparent active motion, creation and annihilation of surface defects in the turbulent regime as provided via effective elastic line tension and splitting of bulk defect lines at the surface. In different realizations, though, we very much envisage that the velocity boundary condition could be different, from no slip, to partial slip or (full) slip, in which case one would expect similar defect dynamics, but now also propelled through the surface. Alternatively, the defect dynamics might be also affected by different surface and  bulk rotational viscosities, effectively, causing the defects in the bulk and at the surface to move at different time scales. Finally, such combined actively-driven surface and bulk  dynamics might lead to the emergence of effectively coupled or decoupled dynamics of defects at the surface and in the bulk.

\section{Discussion}
Confinement is today seen as one the major routes for controlling active matter --- including active nematics --- with multiple geometries explored~\cite{WuK_Science355_2017,WiolandH_NaturePhys12_2016,LushiE_ProcNatlAcadSci111_2014,KeberFC_Science345_2014}. From the perspective of this work, we should emphasize that we purposely chose the regime where the droplet radius $R$ is generally comparable but larger than the active length $\xi_\zeta=\sqrt{L/\zeta}$, i.e. $R/\xi_\zeta \lesssim 35$, and the surface interactions (surface anchoring) are strong, such that the confinement has a profound role. The main effect of confinement with the strong homeotropic boundary condition is that it enforces existence of at least one bulk topological defect (loop) by imposing non net-zero topological charge in the bulk of the droplet, and similarly strong planar anchoring imposes non-zero surface topological charge, thus enforcing existence of surface defects. This confinement-imposed existence of defects in the limit of low activity has a notable effect on the onset of active turbulence as it circumvents the need for spontaneous creation of defect pairs. Instead of a spontaneous defect creation, existing defect or defects get transformed --- e.g. existing defect splits into a defect pair. As a result of this lower energy barrier mechanism, in this work we do not observe events where new defects would emerge from an instability of the homogeneous nematic field and would involve  spontaneous appearance of a single loop with zero topological charge, but novel defect loops are instead created by a breakup from an existing defect loop. In similar manner, one could envisage that some surface anchoring profiles or structures with sharp edges, that would cause strong local distortions of nematic could perform as nucleating regions for the emergence of the active nematic defects. Finally, the topology and geometry of the confining surfaces thus can play an important part in generating distinct active regimes. 



The major body of current work in active nematics is for two-dimensional or quasi-two-dimensional regimes and geometries~\cite{BlanchMercaderC_PhysRevLett120_2018,SanchezT_Nature491_2012}. From the perspective of topological defects, the key difference between 2D and 3D active nematics is that the defects change from defect points in plane to defect lines and loops in three-dimensional space, which introduces a  notable increase in the complexity of tracking and monitoring apporaches for the defects. Topologically, the key difference is that in 2D nematic fields only $+1/2$ and $-1/2$ defects are possible, whereas  in 3D  nematics the defects can be in the from of points, loops or walls, and even lines if these can terminate appropriatelly like on surfaces. Also, solitonic solutions are starting to be seen in objects like active skyrmions~\cite{MetselaarL_JChemPhys150_2019}. These three-dimensional nematic defect structures can be characterised with different topological invariants (in 2D nematics there is only 2D topological charge) that account for the topology of the defects as a whole, such as 3D topological charge, linking and self-linking number. Defects can be characterized also locally through their cross-section, for example with the winding number. At the level of orientational order fields, active nematics are topologically equivalent to the  passive (i.e. not active) nematics, in which 3D defect structures as complex as knots and links were demonstrated~\cite{TkalecU_Science333_2011,MartinezA_NatMater13_2014}. It is an interesting question if similarly complex three-dimensional topological defect structures could emerge also in active nematics and moreover, if they could have some role in real living matter.

In conclusion, using numerical modelling, we demonstrate the topology of three-dimensional active turbulence in droplets with fixed spherical shape of active nematic, considering both droplets with perpendicular and inplane imposed surface orientation. Structural transition with developed hysteresis between bulk regular  offset point defect regime and irregular regime of active turbulence is shown for homeotropic droplets. Defect length, defect position, angular momentum and number of defect loops are used as bulk material characteristics that can account for the coupled material flow, orientation and topological variability of the three-dimension active nematic in turbulent and non-turbulent regime. The bulk active defect lines and loops are shown to exhibit local profiles with spatially and in-time varying cross-section, with winding number of $+1/2$, $-1/2$ and twisted profile. Individually, defect loops are generally topological-charge neutral, except for compensating the charge imposed by the confining surface. Topological flow diagram is shown as a strong tool for topological characterisation of active turbulence, identifying distinct set of topology-affecting or changing events --- defect splitting, annihilation, merging and crossovers --- with their conservation laws drawing on analogy with Morse theory. For droplets with degenerate inplane surface alignment, we show that active defects can touch (wet) the surface, creating active defect line segments that span between different surface regions, topologically coupling the surface and bulk defect dynamics. More generally, the aim of this work is to approach --- in full three-dimensions --- the complex temporal and spatial variability of active matter using a topological tool-box and moreover, to explain the observed continual dynamics in view of distinct transitions between selected well-characterized topological structures.


\section*{Acknowledgments}
Authors acknowledge discussion on 3D active defect lines with D. Beller.
The work was supported by the Slovenian Research Agency (ARRS) under contracts P1-0099, J1-9149 and L1-8135, and COST action EUTOPIA (CA17139). M.R. also acknowledges support under EPSRC grant EP/R014604/1 at Isaac Newton Institute, University of Cambridge (programme The mathematical design of new materials).

\section*{Author contributions}

S. Č., J. A. and Ž. K. contributed equally to the work. J. A. and Ž. K. performed numerical simulations and analysed the results. S. Č. performed geometric and topological analysis. Ž. K. developed numerical code. M. R. and S. Ž. supervised the work. All authors contributed to writing of the manuscript. M.R. initiated and led the research.

\section*{Appendix: Methods}

\subsection*{1. Modelling of active nematics}
We apply mesoscopic continuum description of (dense) active nematic, that is based on coupled dynamic equations for orientational order and the material flow field~\cite{DoostmohammadiA_NatCommun9_2018, HatwalneY_PhysRevLett92_2004, GuillamatP_ScienceAdvances4_2018}. Roughly, this approach  is based on the material flow generation via the active forces caused by the distortions in the orientational ordering of the active nematic, and the back-coupled response of the active nematic orientation to the generated flow. The nematic ordering is described by a traceless tensor order parameter $Q_{ij}$ whose largest eigenvalue is the degree of order $S$ (also called scalar order parameter) and the corresponding eigenvector gives the main ordering axis, called  the director $\mathbf{n}$. The dynamics of the Q-tensor is given by the adapted Beris-Edwards equation~\cite{book-Beris}
\begin{equation}
\left(\partial_t+u_k\partial_k\right)Q_{ij}-S_{ij}=\Gamma^\text{bulk} H_{ij},
\label{eq:BE}
\end{equation}
where $\partial_t$ is a derivative over time $t$, $\partial_k$ is a derivative over $k$-th spatial coordinate, $\vec{u}$ is fluid velocity, and $\Gamma^\text{bulk}$ the rotational viscosity coefficient. Molecular field $H_{ij}$ drives the system towards the equilibrium of $Q_{ij}$ and can be interpreted from the free energy $F$ (but also alternatively, see Ref.~\cite{DoostmohammadiA_NatCommun9_2018}):
\begin{equation}
H_{ij}=-\frac{\delta F}{\delta Q_{ij}}+\frac{\delta_{ij}}{3}\text{Tr}\frac{\delta F}{\delta Q_{kl}},
\end{equation}
where $F$ is written in the Landau-de Gennes form as:
\begin{equation}
\begin{split}
F&=\int_V\Bigl\{\frac{A}{2}Q_{ij}Q_{ji}+\frac{B}{3}Q_{ij}Q_{jk}Q_{ki}+\frac{C}{4}(Q_{ij}Q_{ji})^2 \\
&\phantom{={}}+ \frac{L}{2}\frac{\partial Q_{ij}}{\partial x_k}\frac{\partial Q_{ij}}{\partial x_k}\Bigr\}\mathrm{d}V.\label{eq:F}
\end{split}
\end{equation}
Derivatives of $Q_{ij}$ in Eq.~\ref{eq:F} describe the effective elastic behavior of the director field, where $L$ is the elastic constant. Summation over the repeated indices is implied. $A$, $B$, and $C$ are parameters that can be used to tune the nematic phase behavior.

The advection term $S_{ij}$ couples  the velocity and nematic ordering:
\begin{align}
S_{ij}&=(\chi D_{ik}-\Omega_{ik})\left(Q_{kj}+\frac{\delta_{kj}}{3}\right) \nonumber \\
&\phantom{={}} +\left(Q_{ik}+\frac{\delta_{ik}}{3}\right)(\chi D_{kj}+\Omega_{kj})  \\
&\phantom{={}}-2\chi\left(Q_{ij}+\frac{\delta_{ij}}{3}\right)Q_{kl}W_{lk},\nonumber
\end{align}
where $D_{ij}$ is the symmetric, and $\Omega_{ij}$ the antisymmetric part of the velocity gradient tensor $W_{ij}=\partial_iu_j$. The alignment parameter $\chi$ depends on the molecular shape and defines the flow-aligning or flow-tumbling regime.

Surface alignment is modelled as follows. For droplets with perpendicular surface alignment (homeotropic anchoring), strong anchoring with fixed radial director at the droplet surface is assumed. Droplets with inplane degenerate planar alignment (degenerate planar anchoring) are modelled by the surface free energy density:
\begin{equation}
f_\text{deg}=W_\text{deg}\left(\tilde{Q}_{ij}-\tilde{Q}_{ij}^\perp\right) ^2,
\end{equation}
where $\tilde{Q}_{ij}=Q_{ij}+S\delta_{ij}/2$, $\tilde{Q}_{ij}^\perp=(\delta_{ik}-\nu_i\nu_k)\tilde{Q}_{kl}(\delta_{kj}-\nu_k\nu_j)$, and $\vec\nu$ is the surface normal.
Surface Q-tensor field follows the dynamics
\begin{equation}
\dot{Q}_{ij}=\Gamma^\text{surf}\left( H^\text{surf}_{ij} -\frac{\delta_{ij}}{3}\text{Tr}H_{kl}^\text{surf}\right),
\end{equation}
where
\begin{equation}
H^\text{surf}_{ij}=-\frac{\partial f_\text{vol}}{\partial (\partial_k Q_{ij})}\nu_k-\frac{\partial f_\text{deg}}{\partial Q_{ij}}
\end{equation}
and $f_\text{vol}$ is the free energy density expressed in Eq.~\ref{eq:F}.


The fluid velocity obeys the incompressibility condition and the Navier-Stokes equation
\begin{equation}
\nabla\cdot\vec{u}=0,
\label{eq:continuity}
\end{equation}
\begin{equation}
\rho\left(\partial_t+u_k\partial_k\right)u_i=\partial_j\Pi_{ij},
\label{eq:NS}
\end{equation}
where $\rho$ is the fluid density and $\Pi_{ij}$ the stress tensor, which consists of a passive and an active part $\Pi_{ij}=\Pi_{ij}^\text{passive}+\Pi_{ij}^\text{active}$, where
\begin{align}
\Pi_{ij}^\text{passive}&=-P\delta_{ij}+2\chi\left(Q_{ij}+\frac{\delta_{ij}}{3}\right)Q_{kl}H_{kl} \nonumber \\
&\phantom{={}}-\chi H_{ik}\left(Q_{kj}+\frac{\delta_{kj}}{3}\right) \label{eq:passive} \\
&\phantom{={}}-\chi\left(Q_{ik}+\frac{\delta_{ik}}{3}\right)H_{kj}-\partial_iQ_{kl}\frac{\delta F}{\delta\partial_jQ_{kl}} \nonumber \\
&\phantom{={}}+Q_{ik}H_{kj}-H_{ik}Q_{kj}+2\eta D_{ij},\nonumber \\
\Pi_{ij}^\text{active}&=-\zeta Q_{ij}.
\label{eq:active}
\end{align}
$P$ is fluid pressure, $\eta$ the isotropic viscosity contribution, and $\zeta$ is the activity parameter characterizing the strength of force dipoles of contractile ($\zeta<0$) and extensile ($\zeta>0$) objects.

Coupled equations for the fluid velocity $u_i$ and the nematic order $Q_{ij}$ are solved numerically by the hybrid lattice-Boltzmann algorithm~\cite{DoostmohammadiA_NatCommun8_2017,GuillamatP_ScienceAdvances4_2018}. The hybrid algorithm consists of an explicit finite difference method for the Q-tensor evolution (Eq.~\ref{eq:BE}), and the D3Q19 lattice Boltzmann model for the Navier-Stokes equation and the compressibility condition (Eqs.~\ref{eq:continuity},~\ref{eq:NS}). The nematic stress tensor is implemented in the lattice Boltzmann algorithm as a force contribution with the half-force correction~\cite{book-Kruger}. The spherical droplet cavity is allocated on a rectangular grid and a no-slip boundary condition is implemented by the bounce-back rule. A radial surface normal is allocated on the surface nodes, and it is used for the calculation of the surface Q-tensor field either for homeotropic or for planar degenerate anchoring. We performed multiple test to verify that results are not affected by spurious velocities and numerical method induced symmetry or symmetry breaking. For example, in Fig.~\ref{fig:fig1}a, for different random initial conditions, the defect is displaced from the center in different (arbitrary) directions that are typically not along the (rectangular) simulation grid axes.  The results of the simulations are expressed in the units of elastic constant $L$, nematic correlation length $\xi_\text{N}=\sqrt{L/(A+BS_\text{eq}+\tfrac{9}{2}CS_\text{eq}^2)}$ where $S_\text{eq}$ is the equilibrium nematic degree of order, and nematic intrinsic time scale $\tau_\text{N}=\xi_\text{N}^2/\Gamma^\text{bulk} L$. The phase parameters are set to $A=-0.190\,L/\xi_\text{N}^2$, $B=-2.34\,L/\xi_\text{N}^2$ and $C=1.91\,L/\xi_\text{N}^2$, the nematic is in alignment regime with $\chi=1$, isotropic viscosity contribution equals $\eta=1.38\,\xi_\text{N}^2/L\tau_\text{N}$, and the strength of the planar degenerate anchoring is $W_\text{deg}=6.6\cdot 10^{-4}\, L/\xi_\text{N}$ with surface rotational viscosity parameter $\Gamma^\text{surf}=0.67\,\Gamma^\text{bulk} /\xi_\text{N}$. Grid resolution is set to $\Delta x=1.5\,\xi_\text{N}$ and time resolution to $\Delta t=0.057\,\tau_\text{N}$. Velocity field of the active nematic can be analysed also by calculating the total angular momentum  $\mathbf{\Gamma}=\int \rho \mathbf{r}\times\mathbf{u} \mathrm{d}V$, where $\mathbf{r}$ is the distance from the droplet centre and integration is performed over the whole volume of the active nematic droplet.

The parameters of the considered 3D active nematic system can be summarised by introducing two selected
dimensionless numbers (according to Eqs.~\ref{eq:BE} and \ref{eq:NS}): Ericksen number $\emph{Er}=uR/\Gamma^\text{bulk} L$ comparing the viscous terms to the elastic terms and activity number $\emph{Ac}=\sqrt{\zeta R^2/L}$ characterizing the relative strength of activity vs. confinement, where $u$ is maximum velocity in droplets of radius $R$ and activity $\zeta$. In our simulations we take activities of up to $\zeta\lesssim0.06\,L/\xi_\text{N}^2$ and droplet radius of $R=136\,\xi_\text{N}$ which corresponds to  $\emph{Er}\lesssim 100$ and $\emph{Ac}\lesssim 35$.



\subsection*{2. Tracing and visualisation of defects}
Modelling of active nematics gives a continuous three-dimensional Q-tensor field and a velocity field on a discrete grid at each time step.
To perform the topological analysis, geometry of defect lines is extracted from the Q field. Defects are detected by first finding the point in the sample that has the lowest degree of order $S$.
A small nearby sphere with $3$ data points in radius is searched for the lowest $S$ and the line is propagated until it closes the defect loop by meeting the initial point.
Visited points are removed from the search space, and the algorithm is repeated until no regions with sufficiently low order are found.
This procedure produces polygonal lines that approximate the defect loops, and an associated coordinate frame is constructed, consisting of the tangent $\hat{t}$ and two perpendicular vectors $\hat{t}_{1}$ and $\hat{t}_{2}$. The choice of these two vectors is in principle arbitrary, as long as they vary continuously around the loop. For each defect loop segment, the surrounding director field  is analysed on a small circle spanned by the perpendicular vectors $\hat{t}_1$ and $\hat{t}_2$.
Two mutually perpendicular normalized vectors $\hat{n}_1$ and $\hat{n}_2$ are found which are used to parametrise the circle on the unit sphere of all directions, visited by the director as traversing the circle. These directions $\hat{n}_1$ and $\hat{n}_2$ are flipped head-to-tail if needed so that they vary continuously along the loop.
Keeping the signs and orientation of the local coordinate frame consistent with respect to the initial reference point overcomes the ambiguities inherent in the line field topology.
Note that we trace two coordinate frames over the entire loop. The first frame consists of $\hat{t}$, $\hat{t}_{1}$, and $\hat{t}_{2}$ and  the second frame consists of $\hat{n}_1$, $\hat{n}_2$, and $\hat{n}_1 \times \hat{n}_2$. The orientations of these frames are represented with quaternions instead of rotational matrices, to retain the full topological information in a closed loop~\cite{CoparS_PhysRep538_2014}. The pair $\hat{n}_1$ and $\hat{n}_2$ define the discs, drawn in Fig. 3b to represent the cross section, while the colour represents the quantity $\hat{t}\cdot(\hat{n}_1\times\hat{n}_2)/2$ that distinguishes the winding number, from pure $-1/2$ profile in red, to pure $+1/2$ in blue. Splay-bend parameter is used to visualise local director around the active defect loops~\cite{NikkhouM_NaturePhys11_2015}.

Velocity field in Fig. 1 is shown by blue arrows above selected cut-off magnitudes of      $v=0.0320\,\xi_\text{N}/\tau_\text{N}$ in (a) at $\zeta=0.003\, L/\xi_\text{N}^2$, $v=0.127\,\xi_\text{N}/\tau_\text{N}$ in (a) at $\zeta=0.013\, L/\xi_\text{N}^2$, $v=0.195\,\xi_\text{N}/\tau_\text{N}$ in (a) at $\zeta=0.022\, L/\xi_\text{N}^2$, $v=0.304\,\xi_\text{N}/\tau_\text{N}$ in (b), and $v=0.326\,\xi_\text{N}/\tau_\text{N}$ in (d). Velocity magnitude is additionally represented by a bright green isosurfaces at similar cutoff values of $v=0.0305\,\xi_\text{N}/\tau_\text{N}$ in (a) at $\zeta=0.003\, L/\xi_\text{N}^2$, $v=0.125\,\xi_\text{N}/\tau_\text{N}$ in (a) at $\zeta=0.013\, L/\xi_\text{N}^2$, $v=0.0320\,\xi_\text{N}/\tau_\text{N}$ in (a) at $\zeta=0.192\, L/\xi_\text{N}^2$, $v=0.297\,\xi_\text{N}/\tau_\text{N}$ in (b), and $v=0.316\,\xi_\text{N}/\tau_\text{N}$ in (d).

\section*{Appendix: Supplementary figures}

\begin{figure}[b]
\centering
\includegraphics[width=\textwidth]{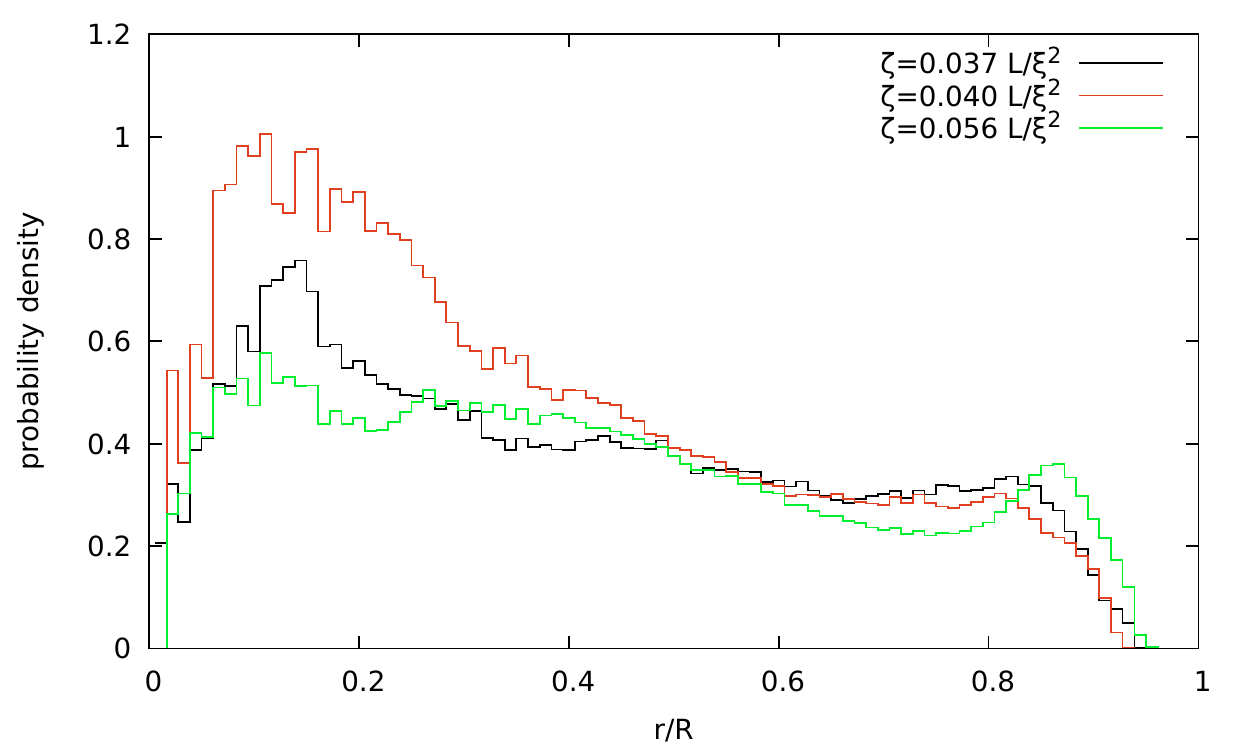}
{SUPPLEMENTARY FIG. 1.
\textbf{Distribution of defect line segments within the droplet.}
Probability density of defect line segments per unit volume, with respect to relative distance from the center of the droplet. The probability is normalized to unity when integrated over the whole volume of the droplet. All three graphs are in the active turbulent regime. At lower activities, the defects are more likely found in the middle of the droplet, whereas with increasing activity the probability becomes more uniform. Immediately adjacent to the surface, probability is somewhat larger due to the flow spreading and extending the defects against the surface.
}
\end{figure}

\begin{figure}
\centering
\includegraphics[width=0.8\textwidth]{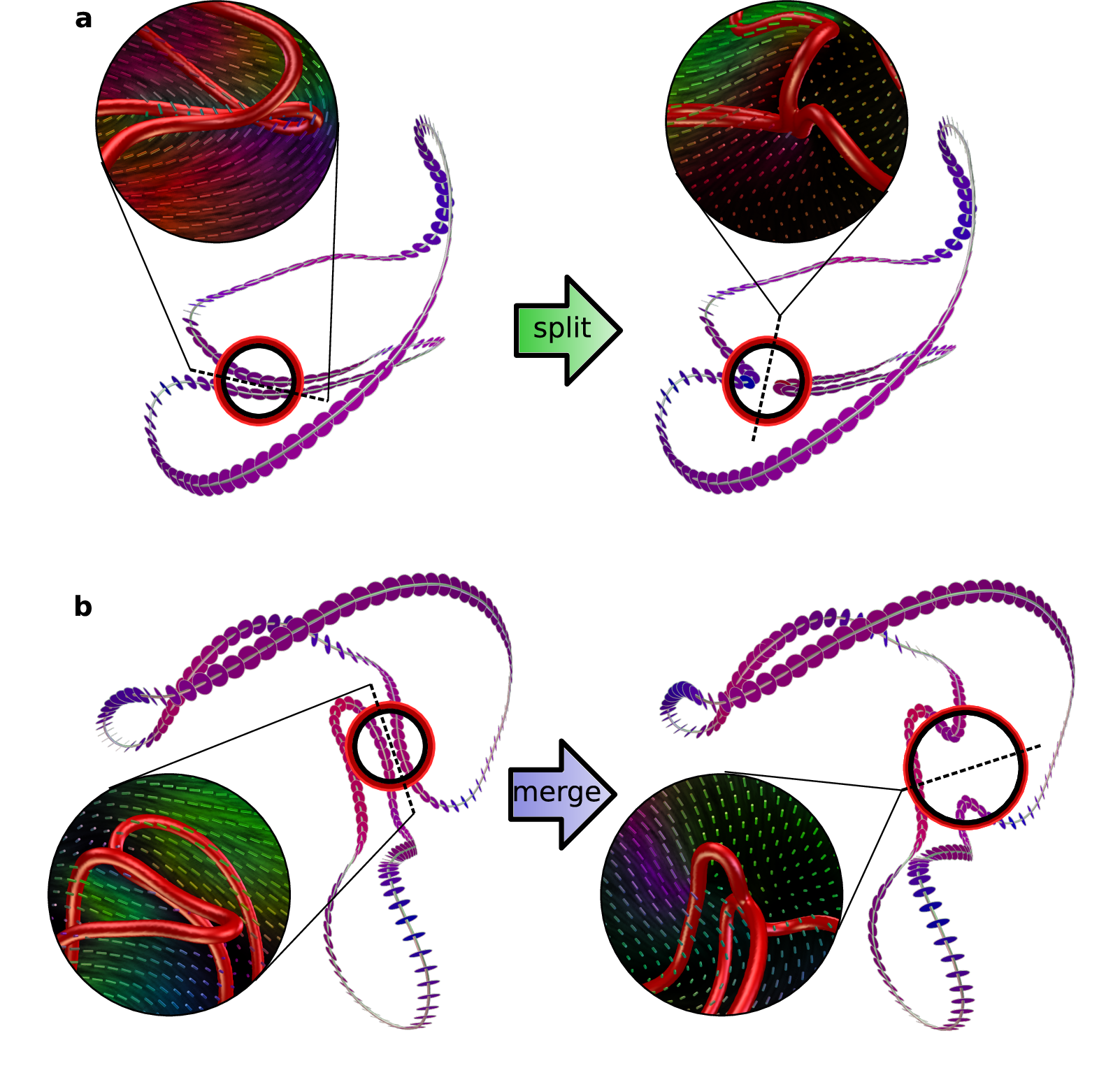}
SUPPLEMENTARY FIG. 2.
\textbf{Defect cross sections during two selected rewiring events.} Splits (a), merges (b) and cross-overs are all based on local rewiring of two defect line segments. The disks showing the plane of directions the director takes around the defect line are shown to be coplanar between the segments both before and after the rewiring.  The cross sections are also topologically complementary – a pair of $-1/2$ and $+1/2$ profiles (red and blue), a pair of oppositely twisted profiles (purple), or a general intermediate pair. The insets show the director in the plane that separates the defect line segments. In both presented events, the director is planar to the separation plane before the rewiring and perpendicular to the separation plane after the rewiring. Note that each panel shows all defects present in a droplet at a given time, so the total topological charge in each panel must equal $+1$ (enclosing droplet is not shown). In panel (a), the smaller split-off loop has zero topological charge. In panel (b), the bottom loop before merging has zero topological charge.
\end{figure}

\begin{figure}
\centering
\includegraphics[width=\textwidth]{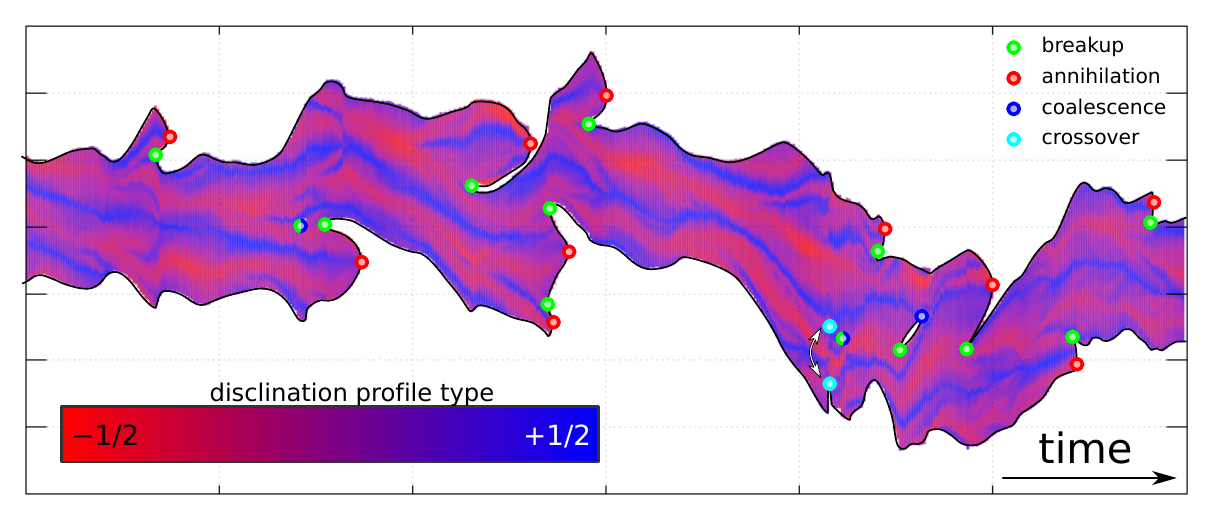}
{SUPPLEMENTARY FIG. 3.
\textbf{Time evolution of defect loop profile in active turbulence regime.} At each time, the defect set is represented by vertical strips with length of each segment roughly corresponding to the defect loop length, and the color corresponding to the director profile variation along the loop. In time series, breakups of defects are seen as branching of a single \quotes{stream} into two, coalescence is the opposite of that, annihilations are apex points, and crossovers are poorly represented in this flat projection.
Note that defects are closed loops, but to represent them as vertical strips, they are cut at an arbitrary position and aligned with those at a previous time step for best matching.
}
\end{figure}

\clearpage

\bibliographystyle{plain}

\bibliography{active_droplets}

\end{document}